\documentclass[10pt,conference]{IEEEtran}
\usepackage[utf8]{inputenc}
\usepackage{amsmath}
\usepackage{xspace}
\usepackage{hyperref}

\newcommand{\linebreakand}{%
  \end{@IEEEauthorhalign}
  \hfill\mbox{}\par
  \mbox{}\hfill\begin{@IEEEauthorhalign}
}

\newcommand{\fb}{\textsc{FuzzBench}\xspace}
\newcommand{\of}{\textsc{OSS-Fuzz}\xspace}
\newcommand{\aflppp}{\textsc{AFL+++}\xspace}
\newcommand{\aflrr}{\textsc{AFL\-rust\-rust}\xspace}
\newcommand{\aflsmart}{\textsc{AFL\-Smart++}\xspace}
\newcommand{\hsfuzz}{\textsc{Haste\-Fuzz}\xspace}
\newcommand{\lpfuzz}{\textsc{Learn\-Perf\-Fuzz}\xspace}
\newcommand{\libaflfuzz}{\textsc{libAFL\_\-lib\-Fuzzer}\xspace}
\newcommand{\pastis}{\textsc{Pastis}\xspace}
\newcommand{\symsan}{\textsc{Symsan}\xspace}
\newcommand{\afl}{\textsc{AFL}\xspace}
\newcommand{\aflpp}{\textsc{AFL++}\xspace}
\newcommand{\hgfuzz}{\textsc{Hongg\-Fuzz}\xspace}
\newcommand{\libfuzzer}{\textsc{lib\-Fuzzer}\xspace}

\title{SBFT Tool Competition 2023 - Fuzzing Track} 
\author{
\IEEEauthorblockN{Dongge Liu\IEEEauthorrefmark{1}}
\IEEEauthorblockA{Google, USA}
\and
\IEEEauthorblockN{Jonathan Metzman\IEEEauthorrefmark{1}}
\IEEEauthorblockA{Google, USA}
\and
\IEEEauthorblockN{Marcel B\"ohme\IEEEauthorrefmark{2}}
\IEEEauthorblockA{MPI-SP, Germany\\Monash, Australia}
\and
\IEEEauthorblockN{Oliver Chang\IEEEauthorrefmark{1}}
\IEEEauthorblockA{Google, USA}
\and
\IEEEauthorblockN{Abhishek Arya\IEEEauthorrefmark{1}}
\IEEEauthorblockA{Google, USA}
\and

\IEEEauthorblockA{
~~~~~~~~~~~~~~~~~~~~~~~~~~~~~~~~~~~~~~~~~\IEEEauthorrefmark{1}\{donggeliu, metzman, ochang, aarya\}@google.com
}
\IEEEauthorblockA{
~~~~~~~~~~~~~~~~~~~~~~~~~~~~~~~~~~~~~~~~~\IEEEauthorrefmark{2} marcel.boehme@mpi-sp.org
}
}

\date{March 2023}

\begin{document}

\maketitle

\begin{abstract}
    This report outlines the objectives, methodology, challenges, and results of the first Fuzzing Competition held at SBFT 2023. The competition utilized \fb to assess the code-coverage performance and bug-finding efficacy of eight participating fuzzers over 23 hours.
    The competition was organized in three phases. In the first phase, participants were asked to integrate their fuzzers into \fb and allowed them to privately run local experiments against the publicly available benchmarks. In the second phase, we publicly ran all submitted fuzzers on the publicly available benchmarks and allowed participants to fix any remaining bugs in their fuzzers. In the third phase, we publicly ran all submitted fuzzers plus three widely-used baseline fuzzers on a hidden set and the publicly available set of benchmark programs to establish the final results.
    
\end{abstract}

\section{Introduction}
We report on the organization of the first fuzzing competition at the 16th International Workshop on Search-Based and Fuzz Testing (SBFT) held on the 14th of May 2023 in Melbourne, Australia. The objectives of this competition were (i)~to evaluate the performance of the fuzzers submitted to this competition in terms of coverage and bug finding ability, (ii)~to gather experience and feedback on the sound benchmarking of fuzzing tools, and (iii)~to stress test the \fb benchmarking platform which has been built particularly for this purpose.

Throughout the competition we paid particular attention to the mitigation of different forms of bias. For instance, in order to avoid overfitting to a particular set of benchmarks (confirmation bias), we allowed participants to develop, integrate, and evaluate their fuzzers privately on a publically available set of benchmarks while conducting the actual competition on a set of benchmarks that included a large number of hidden benchmarks. In order to avoid survivorship bias, we do \emph{not} evaluate their bug finding ability on a \emph{given} set of bugs that we already know how to find. Instead, we evaluate their bug finding ability in terms bugs found by any fuzzer. We make sure to use the same AddressSanitizer (ASAN) instrumented binaries across all fuzzers.

In summary, we found that the \aflrr fuzzer performed well in terms of both, the coverage achieved and bugs found. The fuzzers \libaflfuzz, \hsfuzz, and \aflppp excelled on coverage-based benchmarks, while \pastis and \aflsmart found more bugs than the average fuzzer. We present the final ranking and more concrete results live at the tool competition.

\section{\fb: Fuzzer Benchmarking Platform}
\fb \cite{metzman2021fuzzbench} is a free, open source fuzzer benchmarking service built to make fuzzer benchmarking easy and rigorous. It allows researchers, who are interested in evaluating their fuzzers against other state-of-the-art fuzzers, to launch large-scale experiments in a free and reproducible manner.

The \fb infrastructure consists of a large number of publicly available benchmark programs taken from \of\unskip\footnote{\url{https://google.github.io/oss-fuzz/}}. The benchmark programs are open source C/C++ programs carefully integrated by their maintainers, and include programs like \texttt{Curl}\footnote{\url{https://github.com/curl/curl}}, \texttt{OpenSSL}\footnote{\url{https://github.com/openssl/openssl}}, \texttt{PHP}\footnote{\url{https://github.com/php/php-src}}, and \texttt{systemd}\footnote{\url{https://github.com/systemd/systemd}}.
Because the source code for most \fb experiments is made public and the specific \fb version can be pinned, reproducing \fb experiments is often much easier than reproducing bespoke experiments used in other research.

\fb can conduct bug-based or code coverage-based experiments \cite{asprone2022comparing}.
Throughout out the course of an experiment, and upon its completion, \fb generates a report detailing the performance of each fuzzer.
The report compares fuzzers based on their performance across all benchmarks as well as on individual benchmarks and shows effect size (Vargha Delaney $\hat A_{12}$) and statistical significance (Mann Whitney $U$ test).
The comparison across all benchmarks contains two rankings, one based on their average rank on each individual benchmark and one based on their performance relative to the best performing fuzzer on each individual benchmark.
\fb reports include a critical difference diagram so that users can see if differences between fuzzers based on average rank is statistically significant.
The report's comparison on individual benchmarks consists of graphs and data showing,  the number of crashes found and the growth of code coverage throughout the experiment.

To request an experiment, the interested researcher submits a pull request to the Github repository where the fuzzer is integrated or privately emails fuzzbench@google.com.
A typical experiment in \fb involves about 20 trials of 10 fuzzers running on 20 benchmarks for 23 hours. This is about 10-CPU years, which is cost prohibitive for most researchers.
Researchers can use \fb by integrating with a simple Python and Docker based API.
This integration usually is less than 100 lines of code.


\fb has had an enormous impact on fuzzer development and research. Over 900 experiments have been conducted using the \fb service. \fb has been discussed in over 100 academic papers. And \fb has been used to guide the development of popular fuzzers such as \aflpp, \hgfuzz and \libfuzzer. 
\fb experiments have most desirable qualities that Klee et al. \cite{evaluating_fuzz_testing} described most evaluations as lacking, including: statistically sound comparisons and statistical tests, long timeouts and real-world programs.


\section{Competition Setup}

\textbf{Phases}. The competition was organized in three phases. In the first phase, participants were asked to integrate their fuzzers into \fb and allowed them to privately run local experiments against the publicly available benchmarks. In the second phase, we publicly ran all submitted fuzzers on the publicly available benchmarks and allowed participants to fix any remaining bugs in their fuzzers. In the third phase, we publicly ran all submitted fuzzers plus three widely-used baseline fuzzers on a hidden set and the publicly available set of benchmark programs to establish the final results.

\textbf{Performance metrics}.
In our competition, we measure both the code coverage achieved and the bug-finding capacity to compare the performance of the submitted fuzzers  \cite{evaluating_fuzz_testing, bohme2022reliability}. As benchmarking platform, we use \fb which measures line coverage across all coverage-based benchmarks and the time it takes to generate the first crashing input across all bug-based benchmarks. To facilitate a more intuitive comparison of fuzzer performance in both categories, we present a \textit{relative median score} for each fuzzer.

We compute the coverage-based score for each fuzzer as follows. As it is impractical to determine the total number of reachable lines in each coverage-based benchmark $bc$ \cite{effectiveness}, we compute the relative coverage score $score(bc,f)$ for a fuzzer $f$ by dividing the median value of its line coverage over $20$ trials (i.e., $\operatorname{cov}(bc,f,n)$ where $n=1..20$) by the maximum line coverage attained by all fuzzers $F$ on that specific benchmark:

\begin{equation}
score(bc, f) =
\frac{\operatorname{cov}(bc,f)}
{\max\limits_{i \in F}{\max\limits_{n=1..20} \operatorname{cov}(bc, i, n)}}
\end{equation}
\begin{equation}
\operatorname{cov}(bc,f) = \substack{\operatorname{Med} \\ n=1..20}(\operatorname{cov}(bc, f, n))
\end{equation}

We compute the bug-based score for each fuzzer as follows. Many fuzzer-generated crashing inputs may expose the same bug, and the same bug may yield different stack traces \cite{rebucket, tracesim}. In order to circumvent challenges of bug deduplication, we include only one reproducible bug in each benchmark and measure the time it takes to generate the first input that causes the benchmark binary to crash. Therefore, considering that each bug-based benchmark $bb$ comprises only one bug, we calculate the relative score $score(bb,f)$ of a fuzzer $f$ using the following method:

\begin{equation}
score(bb, f) = \substack{\operatorname{Med} \\ n=1..20}(\operatorname{bug}(bb, f, n)) 
\end{equation}
\begin{equation}
\operatorname{bug}(bb, f, n) = \begin{cases}
      $1$ & \text{if $f$ finds a bug in $bb$ in trial $n$}\\
      $0$ & \text{otherwise}
\end{cases}
\end{equation}

In instances where multiple fuzzers detect an equal number of bugs across all benchmarks, we additionally provide their average time required for bug discovery as an auxiliary metric.

\textbf{Benchmarks.}
The $53$ benchmarks employed in this study were selected from a diverse range of real-world open-source projects integrated into \of.
This approach ensures that researchers can evaluate their fuzzers on the latest, popular, and actively maintained real-world open-source programs. Meanwhile, project maintainers can benefit from state-of-the-art fuzzers.

To guarantee the reproducibility of fuzzer performance, each benchmark is anchored to a specific commit. In particular, the commit for each bug-based benchmark are carefully chosen such that the bug present have been fixed or published within one year. This approach prevents security vulnerability leakage while maintaining benchmarks up-to-date for research evaluation purposes.

Benchmarks are divided into public and private sets. The \emph{public benchmark set}, consisting of 5 bug-based and 24 coverage-based benchmarks, is made available to participants for build and runtime errors identification upon joining the competition. In contrast, the \emph{private benchmark set}, comprising of 10 bug-based and 14 coverage-based benchmarks, is withheld until the final evaluation to mitigate overfitting.

Preventing overfitting in fuzzing competitions is typically challenging since participants usually require access to the benchmark source code to identify and resolve compatibility issues. However, \fb\unskip's design effectively addresses this issue by separating the benchmarks and fuzzers. This allows fuzzers to be built and run on private benchmarks using the same code that was tested on the public ones, contributing to a fair and impartial evaluation of fuzzer performance.

\textbf{Fuzzers.}
The competition evaluates a total of 12 fuzzers, including 8 fuzzers submitted by participants and 4 fuzzers used as baseline. The participant-submitted fuzzers are \aflppp\unskip\footnote{\url{https://github.com/google/fuzzbench/tree/SBFT'23/fuzzers/aflplusplusplus}}, \aflrr\footnote{\url{https://github.com/google/fuzzbench/tree/SBFT'23/fuzzers/aflrustrust}}, \aflsmart\footnote{\url{https://github.com/google/fuzzbench/tree/SBFT'23/fuzzers/aflsmart\_plusplus}}, \hsfuzz\footnote{\url{https://github.com/google/fuzzbench/tree/SBFT'23/fuzzers/hastefuzz}}, \lpfuzz\footnote{\url{https://github.com/google/fuzzbench/tree/SBFT'23/fuzzers/learnperffuzz}}, \libaflfuzz\footnote{\url{https://github.com/google/fuzzbench/tree/SBFT'23/fuzzers/libafl\_libfuzzer}}, \pastis\footnote{\url{https://github.com/google/fuzzbench/tree/SBFT'23/fuzzers/pastis}}, and \symsan\footnote{\url{https://github.com/google/fuzzbench/tree/SBFT'23/fuzzers/symsan}}.
The four baseline fuzzers encompass \afl\footnote{\url{https://github.com/google/fuzzbench/tree/SBFT'23/fuzzers/afl}}, \aflpp\footnote{\url{https://github.com/google/fuzzbench/tree/SBFT'23/fuzzers/aflplusplus}}, \hgfuzz\footnote{\url{https://github.com/google/fuzzbench/tree/SBFT'23/fuzzers/honggfuzz}}, and \libfuzzer\footnote{\url{https://github.com/google/fuzzbench/tree/SBFT'23/fuzzers/libfuzzer}}.
We selected \afl and \aflpp as baselines, as most participants extended them to construct their own.
The fuzzers \hgfuzz and \libfuzzer were chosen due to their contribution to the discovery of bugs in the bug-based benchmarks under \of production environment.


\textbf{Platform and Configuration.}
The competition is conducted on \texttt{Google} Cloud virtual machines.
We concurrently measure 20 trials per fuzzer on each benchmark, with each trial executing one fuzzer instance on one benchmark.
Each trial was run on a dedicated clean \texttt{Ubuntu20.04} virtual machine instance equipped with $1$ vCPU and $3.75$ GB memory.
For some benchmarks, seed corpora were available, mirroring the production environment in \of.

\section{Evaluation Results}
We present and discuss the results of coverage-based and bug-based benchmarking separately. From previous experiments \cite{bohme2022reliability}, we do not expect a strong agreement between rankings established by coverage-based versus bug-based benchmarking, but they each provide important and interesting insights about the capabilities of the fuzzers.

\subsection{Coverage-based Benchmarking}
We first focus on the fuzzers' ability to cover the most code possible. Bugs cannot be found in code that is not covered.

We find that \libaflfuzz leads in 23 out of 38 coverage-based benchmarks, significantly more than any other fuzzer.
However, its overall performance is negatively impacted by the near-zero coverage exhibited on three benchmarks: \texttt{draco}, \texttt{dropbear}, and \texttt{proj4}. In particular, \libaflfuzz generated merely two input cases for \texttt{draco} and crashed immediately after initiating \texttt{dropbear}. To facilitate debugging, \fb has provided researchers with the input corpora and fuzzer logs.

\hsfuzz consistently performs well on all coverage-based benchmarks, securing its position as one of the best fuzzers. Although its relative median scores ranked first on only $16$ benchmarks, it remained within the $90\%$ relative median range on $31$ benchmarks and secured a position within the top three on $35$ benchmarks. Notably, it exhibited the lowest standard deviation across all benchmarks (approximately $8.15$), which is less than half of the second-lowest (\aflppp, $17.61$).

Both \aflppp and \aflrr display competitive performance across the majority of benchmarks. Their relative scores ranked first on $16$ and $12$ benchmarks, respectively, achieved within the $90\%$ range on $31$ and $29$ benchmarks, and secured top three positions on $33$ and $22$ benchmarks.

As for baseline fuzzers, \aflpp emerged as the best-performing and outperformed most other fuzzers on the majority of benchmarks. Its average relative score is $92.67$, whereas the highest average score among the remaining participant fuzzers is below $90$.

A notable observation is that many top-performing fuzzers exhibit a high degree of similarity in their coverage performance, primarily due to their shared underlying fuzzer architecture. To measure "coverage similarity", we consider the coverage achieved by two fuzzers across different benchmarks and compute the cosine similarity. We find that the cosine similarity between \aflrr and \aflppp surpasses $0.99$, signifying their nearly identical relative median scores across all benchmarks. Likewise, the cosine similarities among \aflrr and \hsfuzz, \hsfuzz and \aflpp,  \aflpp and \aflppp are all above $0.98$. In contrast, the cosine similarities between \libfuzzer and \aflpp, \libfuzzer and \aflrr, \libfuzzer and \aflppp are approximately $0.93$.

Our analysis reveals that certain benchmarks are adept at distinguish the coverage performance of fuzzers. For instance, after excluding outliers, the \texttt{openthread} benchmark exhibits the highest interquartile range of $22.25$, along with a standard deviation of $18.91$. The range of fuzzer scores on this benchmark spans from $98$ to $49$, indicating that the top-performing fuzzer achieves approximately double the relative coverage of the lowest-performing one.

Similarly, the scores on the \texttt{lcms} benchmark range from $95$ to $19$, yielding a standard deviation of $22.79$ and an interquartile range of $18.50$. For the \texttt{freetype2} benchmark, the standard deviation is $19.98$, with an interquartile range of $21.75$ and fuzzer scores ranging from $22$ to $95$. Furthermore, no fuzzer's relative median score exceeds $68$ on the \texttt{botan} benchmark, suggesting that the maximum of median scores of all fuzzers is approximately two-thirds of the highest line coverage across all trials.

Conversely, some benchmarks display a high degree of similarity in performance across fuzzers, thereby offering limited utility in differentiating and ranking them. For example, all fuzzers are within $98\%$ of the top-performing fuzzer's score on the \texttt{libjpeg} benchmark, and almost all of them achieve the same line coverage on the \texttt{firestore} benchmark.

\subsection{Bug-based Benchmarking}

In terms of bug finding, many fuzzers display similar performance on bug-based benchmarks. For instance, \aflrr and \pastis both have the highest relative median score ($53.33$), indicating that their median-performing fuzzer trials covered $8$ out of $15$ bugs across all benchmarks. Likewise, participant-submitted fuzzers \aflppp and \hsfuzz covered $6$ bugs, equal to the performance of baseline fuzzers \aflpp and \libfuzzer.

Seven benchmarks were found to be particularly useful in differentiating fuzzers in this competition, as they exhibited diverse performance among fuzzers: \texttt{aspell}, \texttt{assimp}, \texttt{file}, \texttt{bloaty}, \texttt{ffmpeg}, \texttt{libaom}, and \texttt{libxml2}. Both \aflrr and \pastis discovered $6$ out of the $7$ bugs in these benchmarks, outperforming other fuzzers. However, \aflrr and \pastis had slightly different bug-finding patterns; \aflrr covered a comparatively rare bug in \texttt{file} but missed a more commonly found bug in \texttt{ffmpeg}.

While half of the fuzzers found more than $4$ bugs overall, the symbolic-based fuzzer \symsan discovered only $1$ bug in \texttt{assimp}. Interestingly, \libaflfuzz, which performed well across coverage-based benchmarks and found bugs in $5$ benchmarks, was the only fuzzer that missed the bug in \texttt{assimp}. This result could be attributed to its relatively low coverage on this specific benchmark.

We also examined the average time required for fuzzers to discover a bug. \pastis proved to be the fastest in detecting bugs on average, with \aflrr and \aflsmart following closely behind. Notably, the cosine similarity between \aflpp and \aflppp exceeds $0.98$, suggesting that they frequently identify bugs at approximately the same time. Likewise, the cosine similarity between \hgfuzz and \pastis surpasses $0.9$, indicating a comparable speed in causing crashes within the benchmark. \libaflfuzz appears to possess a distinct design, resulting in the lowest similarity score when compared to any other fuzzers.

The bug-based benchmarks in this competition also underscore the "asymmetry" between coverage-based and bug-based rankings, as highlighted by B{\"o}hme et al. \cite{bohme2022reliability}. For instance, \hsfuzz excelled in coverage-based benchmarks yet discovered fewer bugs. Conversely, \afl identified more bugs than \aflpp, despite covering less code. Although code coverage is a well-established and easily measurable benchmarking metric, these findings stress the significance of taking bug-finding capabilities into consideration when optimizing for higher coverage and evaluating fuzzers. Essentially, fuzzers are intended to detect bugs, with coverage serving as a heuristic to estimate their bug-finding potential.

Bug-based benchmarking presents several challenges that we tackled in different ways. Firstly, acquiring the source code of real-world bugs is arduous, and the performance measured by artificial bugs might not accurately reflect reality. \fb addresses this issue by using bugs filed by \of when fuzzing actual open-source projects, providing a ground truth for bugs that had been and need to be discovered in production.

Secondly, a systematic approach for selecting appropriate bug benchmarks for evaluation remains absent. For instance, if all fuzzers exhibit similar performance on certain benchmarks, those bugs offer limited value into fuzzer assessment. To mitigate this concern, we incorporated benchmarks that were hidden during development and only revealing during final evaluation, culminating in nine benchmarks that demonstrate varying bug-discovery performances among fuzzers in this competition.

Thirdly, determining the superior fuzzer performance becomes difficult when multiple fuzzers can discover the same bug. To address this, we employ an auxiliary metric, i.e., measuing the average time required by each fuzzer to discover a bug. While \fb evaluates this metric at 15-minute intervals, which may occasionally compromise accuracy, we highlight that this potential risk does not unfairly benefit any specific fuzzer.

Finally, ascertaining whether multiple crashes correspond to the same bug by grouping backtraces poses a considerable challenge.  To tackle this issue, the competition restricts each benchmark to include only one known bug. Each associated open-source project  is subjected to rigorous testing using multiple fuzzers over an extended period to minimize the likelihood of multiple reproducible bugs coexisting within a single benchmark.

\section{Conclusion and Future Work}
In this competition, \fb evaluates participant fuzzers and common baselines, comparing them using a variety of statistical tools. The assessment encompasses two key metrics: code coverage and bug-finding. Benchmarks for both metrics are derived from real-world open-source projects, and all fuzzers are tested under uniform production-like environment.

Moving forward, \fb aims to enhance the statistical analysis by providing more detailed information, particularly concerning lines or bugs that fuzzers failed to cover. Additionally, \fb plans to incorporate a larger collection of bug-based benchmarks to facilitate more comprehensive statistical reasoning.

\bibliographystyle{IEEEtran}
\bibliography{main.bib}
\end{document}